\documentclass[11pt]{article}

\usepackage[utf8]{inputenc}
\usepackage[T1]{fontenc}
\usepackage{palatino}
\usepackage{amsfonts}
\usepackage[margin=1in]{geometry}
\usepackage{etoolbox}
\usepackage{amsmath}
\usepackage{graphicx}
\usepackage{mdframed}
\usepackage{natbib}
\usepackage{hyperref}
\usepackage{booktabs}

\usepackage{authblk}
\author[1]{Douglas Brinkerhoff}
\author[2]{Elizabeth Fischer}
\affil[1]{Department of Computer Science, University of Montana}
\affil[2]{International Arctic Research Center, University of Alaska}

\title{Conditional Flow Matching for Probabilistic Downscaling of Maximum 3-day Snowfall in Alaska}

\begin{document}

\maketitle

\begin{abstract}
Precipitation in complex terrain is governed by orographic processes operating at scales of a few kilometers, yet climate models typically run at resolutions of 50--100~km where this topographic detail is absent. Dynamical downscaling with high-resolution regional models such as WRF can resolve these processes, but the computational cost---months of wall-clock time per scenario---precludes the large ensembles needed for uncertainty quantification. We present WxFlow, a conditional generative model based on flow matching that learns to map coarse-resolution climate model output and high-resolution topography to calibrated probabilistic ensembles of fine-scale precipitation fields. Applied to 4~km WRF simulations of maximum 3-day snowfall over southeast Alaska, WxFlow achieves 87.8\% improvement in spectral fidelity and dramatically lower Continuous Ranked Probability Scores relative to conventional lapse-rate-corrected bicubic downscaling, while generating 50-member ensembles in seconds on a laptop. Ensemble spread is spatially coherent and governed by topography, reflecting physically plausible uncertainty structure. All code is available at \url{https://github.com/glide-ism/wrf-flow}.
\end{abstract}

\section{Problem and Overview}

Mountain precipitation is a critical input for climate adaptation planning, water resource management, and avalanche hazard assessment. However, climate models operate at $\sim$100~km resolution and fail to represent orographic precipitation in complex terrain. Dynamical downscaling via regional climate models such as the Weather Research and Forecasting (WRF) model \citep{skamarock2019} can capture these fine-scale processes, but the computational cost---often requiring months of wall-clock time per climate scenario---severely limits the number of scenarios that can be evaluated in decision-support applications.

We develop WxFlow, a conditional generative model that learns to map coarse-resolution climate model output and high-resolution topographic information to ensembles of plausible high-resolution precipitation fields. WxFlow uses conditional flow matching \citep[CFM;][]{lipman2023}, a framework for training generative models by learning velocity fields that interpolate between noise and data distributions.  By training a neural network velocity field via conditional flow matching, we enable rapid generation of probabilistic ensembles of downscaled precipitation in seconds rather than months, thereby enabling uncertainty quantification for climate adaptation at computationally tractable cost.  

\section{Approach}

\begin{figure}
\centering
\includegraphics[width=0.95\textwidth]{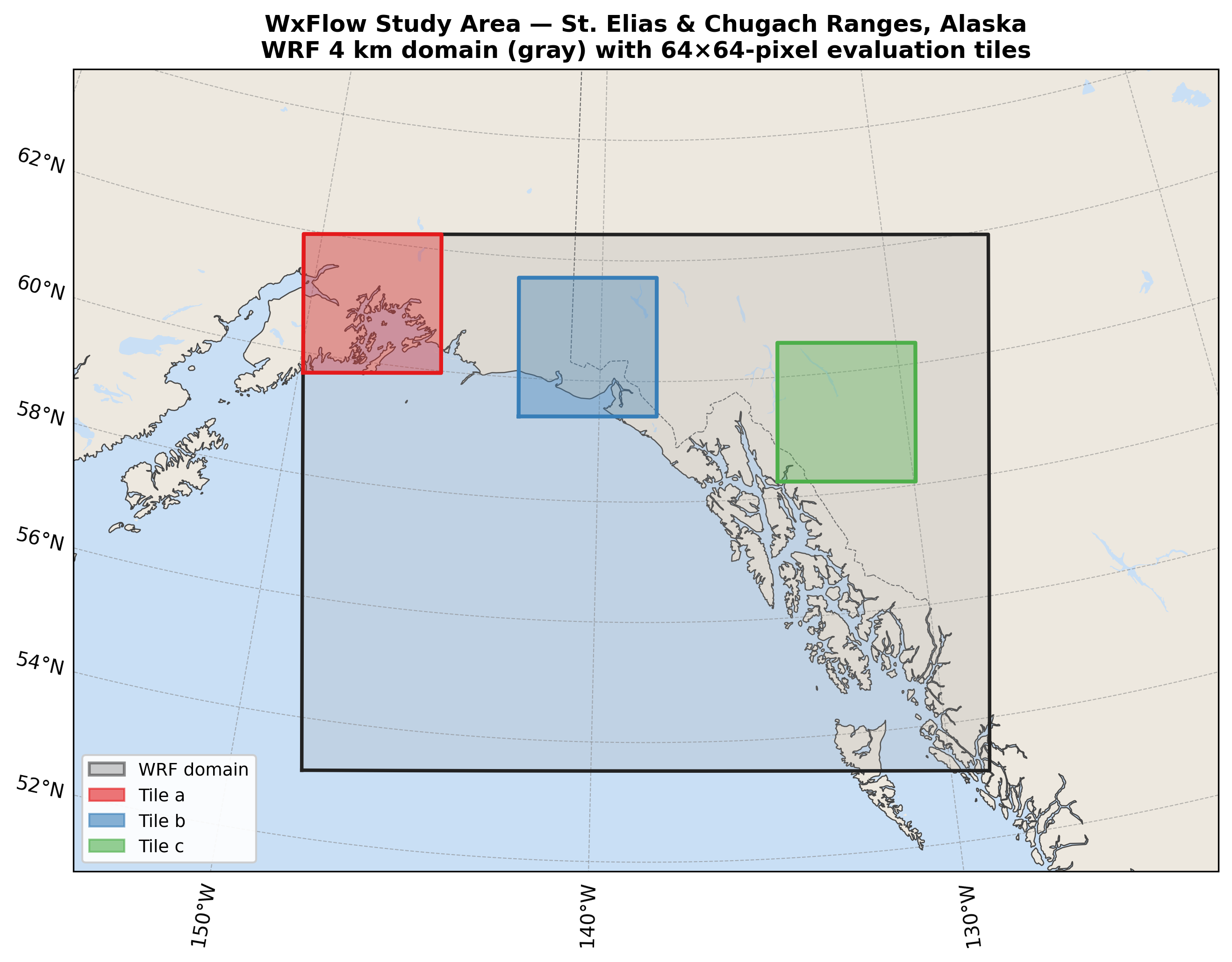}
\caption{Reference map showing regions in Alaska where WxFlow was applied.  Tiles correspond to locations shown in Figs.~2--5}
\label{fig1}
\end{figure}
\begin{figure}
\centering
\includegraphics{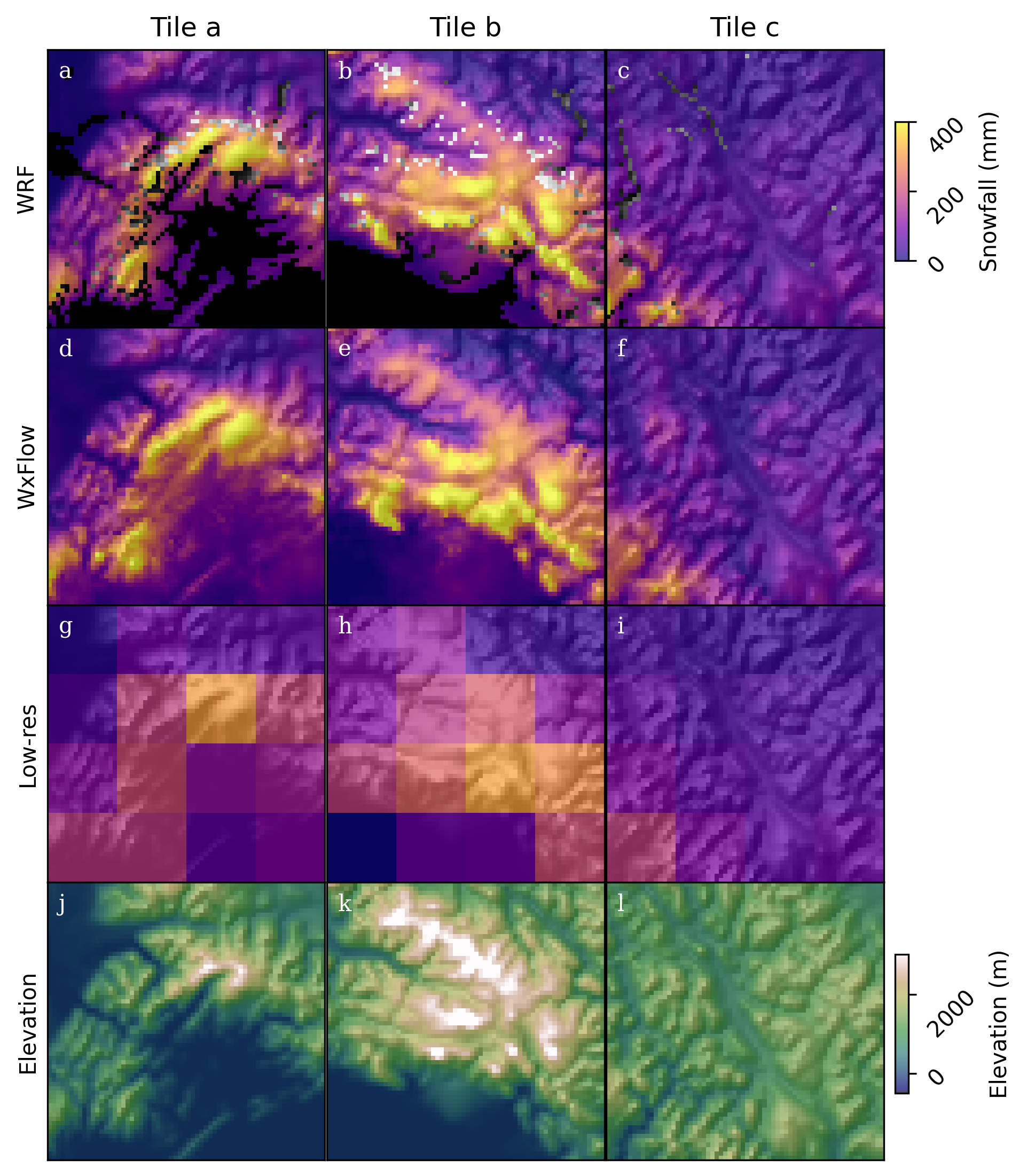}
\caption{Samples drawn from the CFM-based distribution of 3 day maximum snowfall.  The hi-res WRF output (a--c) was coarsened to produce a low-res input for WxFlow (g--i). WxFlow combined the elevation map and low-res input to produce a high-res output (d--f), which shares many qualitative and spectral properties with the physics model.}
\label{fig2}
\end{figure}

To demonstrate its efficacy, we applied WxFlow to the regions of coastal Alaska shown in Fig.~\ref{fig1}.  Beginning with WRF simulation outputs at 4~km horizontal resolution \citep{lader2020}, we coarsened data to simulate 64~km resolution climate model output.  We partitioned the temporal domain into 120 time steps for training and 30+ independent time steps for evaluation.  We then used standard flow matching (Appendix~\ref{Sec:appendix}) to create a generative model of these high-resolution outputs conditioned on topography and low-resolution climate.  The velocity network has approximately 23M parameters and operates on $64\times64$ pixel tiles at 4~km resolution ($256\times256$~km).  The dataset comprises maximum 3-day snowfall accumulations from multiple atmospheric forcings: CFSR reanalysis (1981--2010) for the historical baseline and GFDL CM3 and CCSM4 climate models (2031--2060) for future projections.  Initial results, which took a few seconds to generate on a laptop computer with NVIDIA RTX4070, are shown in Figure~\ref{fig2}.

\section{Ensemble Statistics and Probabilistic Predictions}

\begin{figure}
\centering
\includegraphics[width=\textwidth]{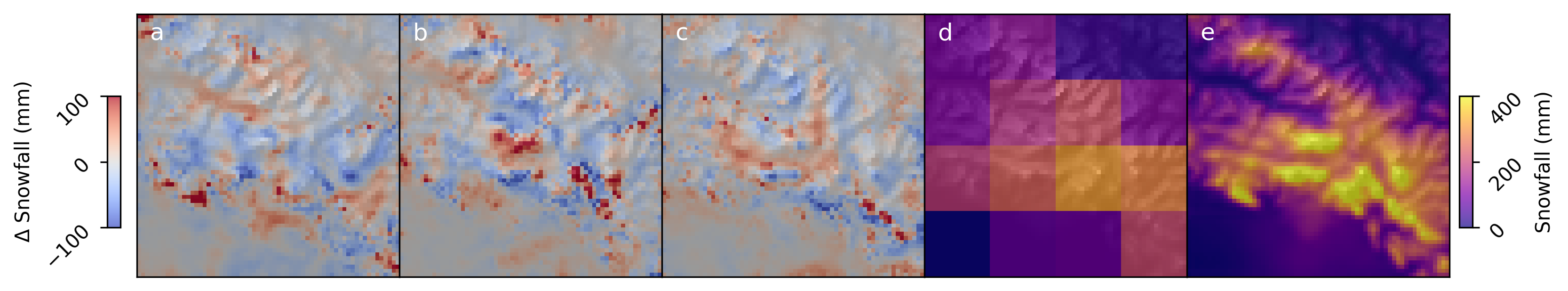}
	\caption{Samples drawn from WxFlow for tile B (Malaspina Glacier area) with ensemble mean removedr (a--c) alongside the low-resolution forcing (d) and ensemble mean (e).  Deviations from the mean are spatially coherent, reflecting topographic effects such as rain-shadowing and orographic enhancement.}
\label{fig3}
\end{figure}
\begin{figure}
\centering
\includegraphics[width=\textwidth]{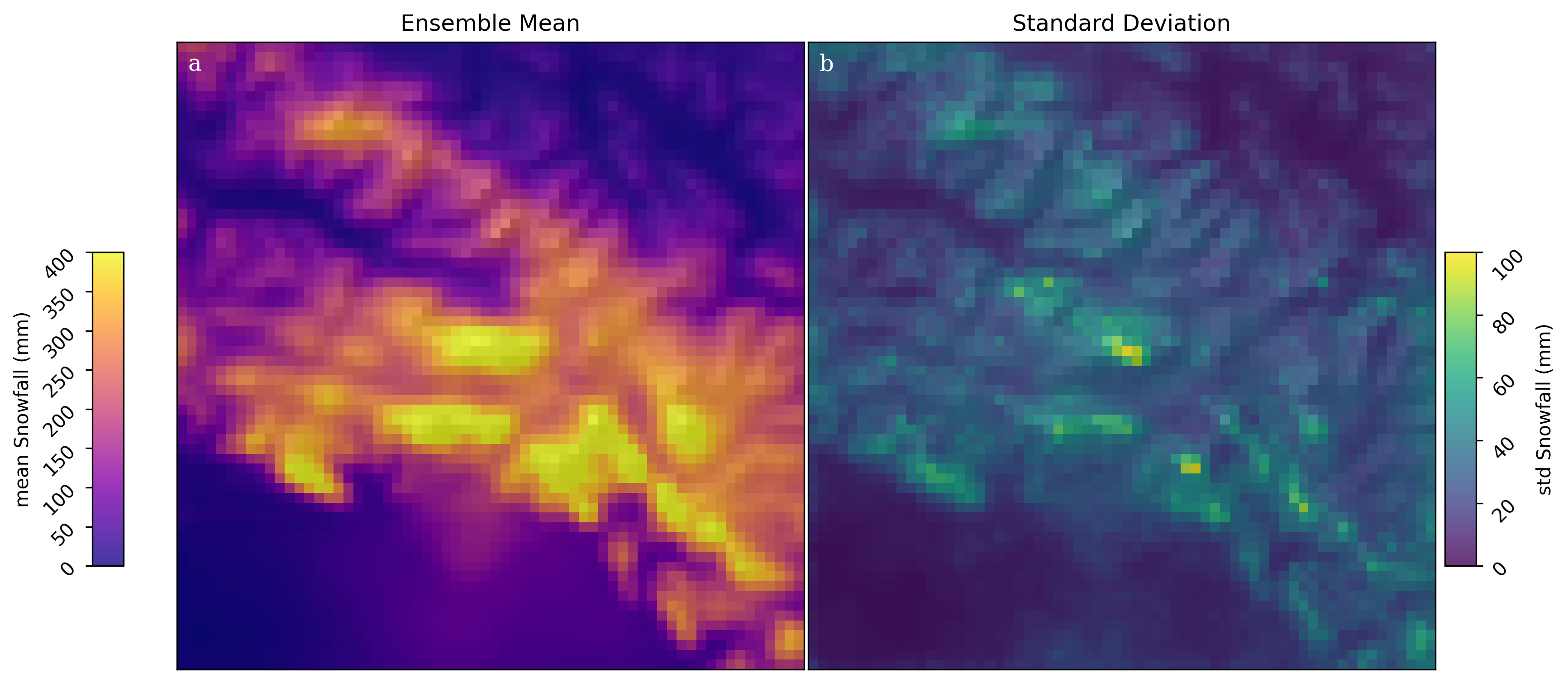}
	\caption{Uncertainty quantification for the St.\ Elias mountain range as quantified by sample standard deviation, centered on tile B (Malaspina Glacier).}
\label{fig4}
\end{figure}
By generating multiple independent realizations with different noise samples, we obtain an ensemble that characterizes the conditional probability distribution over the downscaling problem (Figure~\ref{fig3}). The spatial pattern of ensemble spread provides diagnostic information about which regions have high predictability (low spread) and which regions have intrinsic unpredictability relative to the coarse forcing (high spread).  In particular, WxFlow produces spatially coherent samples in which precipitation anomalies on lee and stoss sides of mountain ridges are correlated across the ridge, reflecting physically plausible rain-shadow effects.  Ensembles can also be characterized by their summary statistics (Fig.~\ref{fig4}), which show that regions of extreme topography and precipitation have greater standard deviation than lower, flatter regions.

\section{Spatial Forecast Skill}
\begin{figure}
\centering
\includegraphics[width=\textwidth]{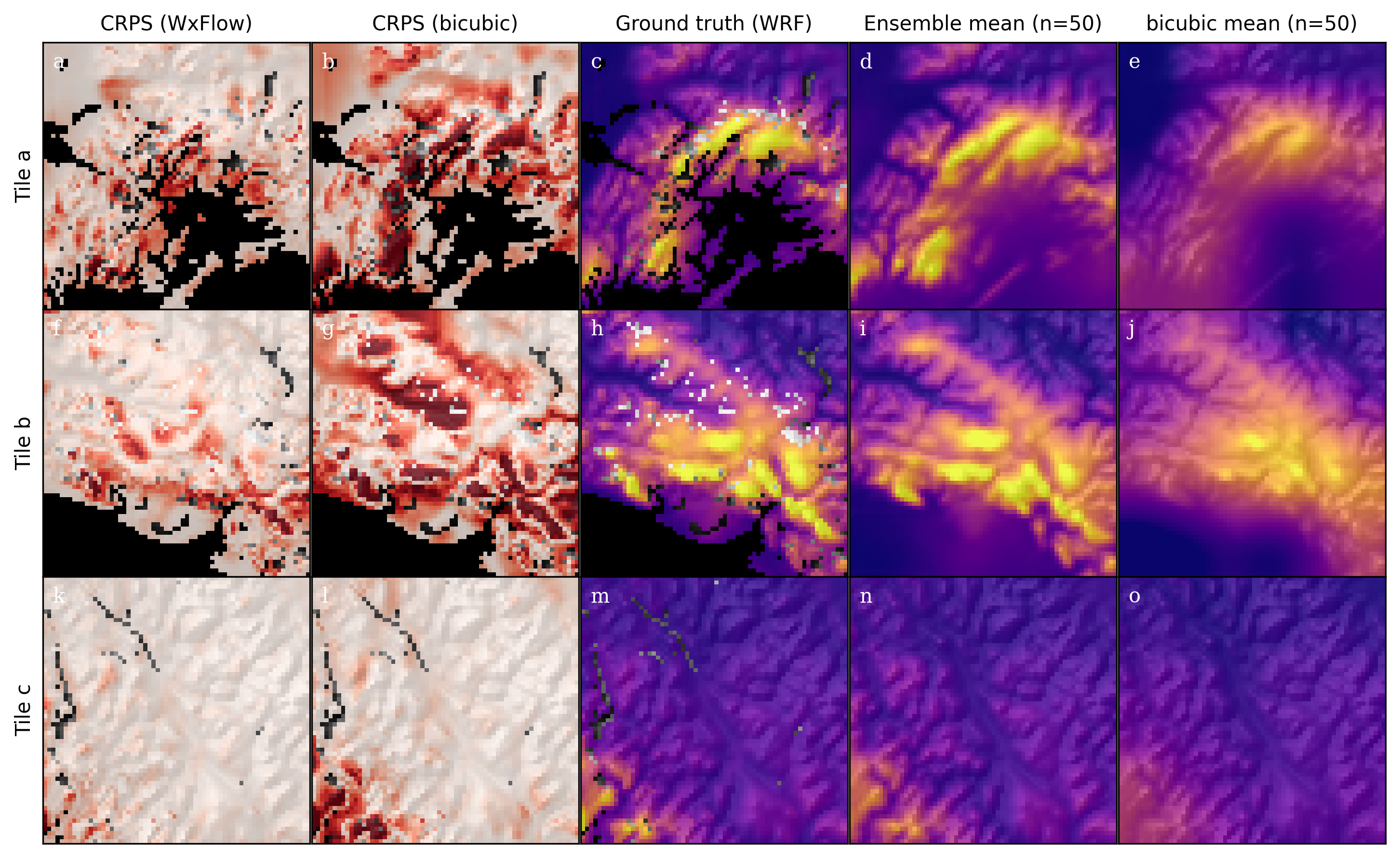}
	\caption{CRPS computed with WxFlow (a, f, k) and the lapse-rate corrected bicubic baseline method (b, g, l) over Tiles a--c in Fig.~\ref{fig1} drawn from the test set, alongside the WRF ground truth (c, h, m), ensemble mean (d, i, n), and baseline mean (e, j, o).  Fields are overlaid on high-resolution topography.  WxFlow exhibits better performance than the baseline with respect to CRPS, particularly over high-relief regions.}
\label{fig5}
\end{figure}

To evaluate WxFlow, we compare it to the baseline bicubic lapse rate method, in which coarse climate data are refined via lapse rate correction.  We used Continuous Ranked Probability Score \citep[CRPS;][]{gneiting2007} to evaluate each of the two methods against the original WRF output.  CRPS is a proper scoring rule that measures the calibration of probabilistic predictions, penalizing both forecast bias and overconfidence while balancing distance to the observation (reliability) against ensemble spread (resolution). A lower CRPS score indicates better forecast skill.  Results (Figure~\ref{fig5}) show that WxFlow dramatically improves accurate placement of precipitation in Southeast Alaska, where orographic precipitation effects are important.

\section{Power Spectral Density and Spatial Structure Fidelity}
\begin{figure}
\centering
\includegraphics[width=\textwidth]{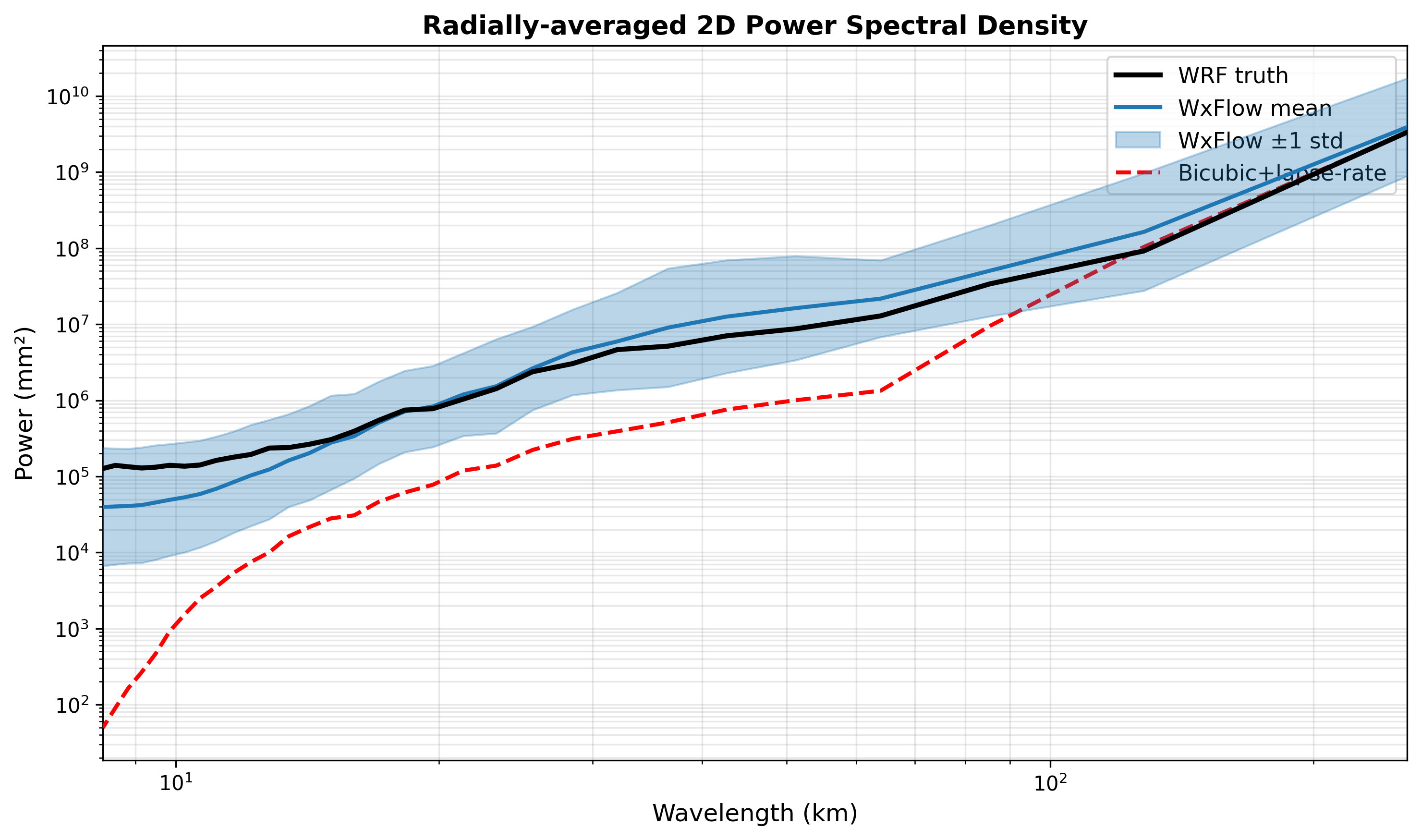}
\caption{Log-log plot of Power Spectral Density for WxFlow and the baseline bicubic algorithm.  WxFlow shows a marked improvement over the baseline, with a little more challenge at high frequencies.  This is consistent with most generative models, which tend to have an easier time learning smooth features.}
\label{fig6}
\end{figure}
While CRPS assesses point-wise forecast calibration, the spatial structure of downscaled precipitation is equally important for applications in hydrology and hazard assessment. We evaluate spatial structure fidelity via the radially-averaged two-dimensional power spectral density (PSD), which decomposes variance as a function of spatial wavelength.
        
The PSD reveals a scale-dependent variance cascade (Figure~\ref{fig6}). Precipitation fields that are over-smoothed exhibit a deficit of power at small scales (short wavelengths $\lambda < 8$ km), while fields with spurious noise show a spectral excess. The WxFlow ensemble span (one standard deviation across 50 members) reveals the uncertainty inherent in the probabilistic prediction at each scale---narrow bands indicate confident predictions, while wide bands indicate high model uncertainty.  

\begin{table}[h]
\centering
\begin{tabular}{lcc}
\toprule
& WxFlow & Bicubic baseline \\
\midrule
Mean abs.\ log spectral error & 0.141 ($\times$1.38) & 1.152 ($\times$14.18) \\
Spectral bias (signed) & $-$0.065 ($\times$0.86) & $-$1.148 ($\times$0.07) \\
\bottomrule
\end{tabular}
\caption{Spectral evaluation metrics (lower magnitude is better).  WxFlow achieves 87.8\% improvement in spectral fidelity over the baseline. Multiplicative factors in parentheses indicate the ratio of model power to truth power implied by the log-space error.}
\label{tab:metrics}
\end{table}

We also evaluate the PSD via two scalar metrics: \emph{mean absolute log spectral error} and \emph{spectral bias} (Table~\ref{tab:metrics}).  Under both, WxFlow significantly improves over the bicubic baseline.  Evaluating the plot by wavelength, we find that WxFlow reproduces WRF's power spectra well over a broad range of wavelengths.  We note that WxFlow exhibits a small ($<$0.3~dB) spectral deficit at high frequencies (on the order of 1--3 pixels; Figure~\ref{fig6}).  Such a result is consistent with most generative models, which tend to have an easier time learning smooth features.  Correcting this systematic bias is an area of ongoing research.

\section{Conclusion}
We have applied conditional flow matching to emulating the annual maximum 3-day snowfall rates in the mountains of coastal Alaska as predicted by WRF, conditioned on high-resolution topography and low-resolution climate forcing similar to that produced by many CMIP-class climate models. We find that WxFlow reproduces high-resolution simulations with high fidelity. The distribution over its predictions is spatially coherent, with deviations from the ensemble mean governed by topography and physically plausible effects like rain shadowing that are evident in the training data. Under quantitative metrics like CRPS and various norms of power spectral density, WxFlow improves significantly over a conventional regression-based baseline. Notably, WxFlow produces spectral consistency with WRF, while exhibiting a modest high-frequency deficit common to generative models. Based on this evidence and previous work on generative downscaling of weather models \citep{mardani2025,pathak2024}, we conclude that conditional flow matching is a promising framework for producing probabilistic, high-resolution precipitation fields as surrogates for computationally expensive dynamical downscaling.  All code required to reproduce these results is available at \url{https://github.com/glide-ism/wrf-flow}.

\appendix
\section{Method Details}
\label{Sec:appendix}

Conditional flow matching \citep{lipman2023} is a framework for training generative models by learning velocity fields that interpolate between noise and data distributions. Unlike diffusion models \citep{ho2020}, which reverse a fixed Markovian noise schedule, flow matching constructs probability paths via optimal transport \citep{chen2018}, enabling more efficient training and faster sampling.

\subsection{Mathematical Framework}
Given a high-resolution precipitation field $\mathbf{x}_1 \sim p_{\mbox{\tiny data}}$ and Gaussian noise $\mathbf{x}_0 \sim \mathcal{N}(0, \mathbf{I})$, the optimal transport probability path interpolates between these distributions:
$$\mathbf{x}_t = \sigma(t) \mathbf{x}_0 + \mu(t) \mathbf{x}_1$$
where $\sigma(t)$ and $\mu(t)$ are smooth monotonic functions satisfying $\sigma(0)=1, \mu(0)=0$ and $\sigma(1)=0, \mu(1)=1$. The velocity field in probability space is then:
$$\mathbf{v}(\mathbf{x}_t, t \mid \mathbf{c}) = \dot{\mu}(t) \mathbf{x}_1 - \dot{\sigma}(t) \mathbf{x}_0$$
where $\mathbf{c}$ denotes the conditioning information (coarse climate and topography). To train the generative model, we parameterize the velocity field $\mathbf{v}_\theta(\mathbf{x}_t, t \mid \mathbf{c})$ via a neural network and minimize the mean squared velocity prediction error:
\begin{equation}
\label{eq:loss}
\mathcal{L} = \mathbb{E}_{t, \mathbf{x}_1, \mathbf{x}_0, \mathbf{c}} \left\| \mathbf{v}_\theta(\mathbf{x}_t, t \mid \mathbf{c}) - \mathbf{v}(\mathbf{x}_t, t \mid \mathbf{c}) \right\|_2^2,
\end{equation}
which is masked to exclude pixels with missing precipitation data. This objective directly minimizes the distance between predicted and target velocity fields, providing a more stable optimization target than score matching used in diffusion models.

\subsection{Architecture: Conditional U-Net with Self-Attention}
    
We employ a U-Net architecture \citep{ronneberger2015} with hierarchical feature extraction and self-attention \citep{vaswani2017} to approximate the conditional velocity field $\mathbf{v}_\theta$. The network has approximately 23M parameters and is designed to balance computational efficiency with expressiveness sufficient to capture multi-scale precipitation patterns.
    
The architecture operates on 4-channel input concatenating (1) the partially denoised precipitation $\mathbf{x}_t$, (2) the high-resolution topography $z$, (3) the coarse-resolution climate model output $\bar{\mathbf{x}}_c$, and (4) a binary validity mask indicating missing data regions. The output is a single-channel velocity field prediction.

The forward pass follows the standard U-Net pattern: an initial convolution expands the spatial resolution to the base channel dimension, a sequence of downsampling blocks progressively reduce spatial resolution while increasing feature depth, a bottleneck layer applies processing at the coarsest resolution, and symmetric upsampling blocks with skip connections restore spatial resolution while progressively decreasing channels.

\paragraph{Time conditioning:} Sinusoidal embeddings of the time variable $t$ are computed via:
\begin{equation}
\mathbf{e}_t^{(j)} =
  \begin{cases}
    \sin(10000^{-2j/d} \cdot t)    & \mbox{if } j \mbox{ is even} \\
    \cos(10000^{-2(j-1)/d} \cdot t) & \mbox{if } j \mbox{ is odd}
  \end{cases}
\end{equation}
These embeddings are projected via an MLP and added to feature maps at each ResBlock, allowing the network to modulate feature processing as a function of the denoising progress.

\paragraph{Multi-scale Attention:} Self-attention layers are applied at $16\times16$ resolution (intermediate depth of the U-Net hierarchy), enabling the network to capture long-range spatial dependencies in precipitation patterns. This resolution balances the computational cost of attention operations with the need to model large-scale orographic forcing.

\paragraph{Normalization and Regularization:} Group normalization (with 32 groups) stabilizes training across variable batch sizes and provides robustness to initialization. Dropout is applied during training to regularize the network and reduce overfitting, particularly important given the modest dataset size.

\subsection{Training Procedure}
    
We train the velocity network via gradient descent on the velocity prediction loss (Eq.~\ref{eq:loss}) over mini-batches of (coarse, fine, topography) triplets sampled from the training set. 

\paragraph{Optimization:} Adam optimizer with learning rate $2 \times 10^{-4}$, $\beta_1 = 0.9$, $\beta_2 = 0.999$.
    
\paragraph{Data Sampling:} Mini-batches of size 24 are randomly sampled from the 120 training time steps, with random spatial tiling applied at each iteration. To counteract the rapid initial signal decay early in the denoising trajectory, the time variable $t$ is sampled with bias toward the latter portion of the [0, 1] interval, increasing the relative importance of the refined precipitation features that emerge as $t \to 1$.
    
\paragraph{Training Duration:} 10,000 optimization steps, corresponding to approximately 2.5 epochs through the training dataset given the random tiling strategy.
    
\paragraph{Masked Loss:} The loss is computed only over pixels with valid (non-missing) precipitation observations, preventing the network from learning to predict velocity in regions where ground truth is unavailable. Checkpoints are saved every 100 steps to enable recovery from training interruption and model selection based on validation loss.
    

\subsection{Generation via ODE Integration}
    
Given a trained velocity network $\mathbf{v}_\theta$, we generate new samples by integrating an ordinary differential equation in probability space. Sampling from noise $\mathbf{x}_0(t=0) \sim \mathcal{N}(0, \mathbf{I})$ and integrating the learned velocity field:
$$\frac{d\mathbf{x}}{dt} = \mathbf{v}_\theta(\mathbf{x}(t), t \mid \mathbf{c})$$
from $t=0$ to $t=1$ with fixed conditioning $\mathbf{c}$ (coarse precipitation and topography), we obtain a sample $\mathbf{x}_1 = \mathbf{x}(t=1)$ from the learned conditional distribution. The ODE is integrated using an adaptive-step Runge-Kutta solver \citep[DOPRI5;][]{dormand1980} with relative and absolute tolerances of $10^{-5}$, requiring approximately 10 function evaluations per sample.

\subsection{Computation and Evaluation of the Power Spectral Density}
To evaluate spatial structure fidelity, we compute the radially-averaged two-dimensional power spectral density (PSD), which decomposes variance as a function of spatial wavelength.  The two-dimensional discrete Fourier transform (FFT) of a precipitation field $\mathbf{x}(i, j)$ on an $n \times n$ grid yields wavenumber components. The power spectral density is defined as:
$$S(k) = \left|\hat{\mathbf{x}}(k_x, k_y)\right|^2$$
where $\hat{\mathbf{x}}$ denotes the 2D FFT and $k = \sqrt{k_x^2 + k_y^2}$ is the radial wavenumber magnitude. To reduce spectral leakage, we apply a 2D Hann window before the FFT. The radially-averaged spectrum is then obtained by binning power into annular wavenumber bands:
$$
\tilde{S}(k) = \frac{\sum_{|\mathbf{k}| = k} S(\mathbf{k})}{\sum_{|\mathbf{k}| = k} 1}
$$
Converting wavenumber to wavelength via $\lambda = 2\pi n \Delta x / k$ (with grid spacing $\Delta x = 4$ km), the spectrum can be plotted on a log-log scale to visualize multi-scale variance structure.

The PSD is evaluated via two metrics, mean absolute log-spectral error
$$
\langle |\log_{10}(S_{\text{model}}) - \log_{10}(S_{\text{truth}})| \rangle$$
averaged over all resolved wavenumbers, quantifying overall spectral fidelity; and spectral bias
$$\langle \log_{10}(S_{\text{model}}) - \log_{10}(S_{\text{truth}}) \rangle,$$
indicating systematic over- or under-prediction of power.

\bibliographystyle{apalike}
\bibliography{bibliography}
\end{document}